\documentclass[aps,prl,superscriptaddress,twocolumn]{revtex4-1}

\usepackage{graphics}
\usepackage{graphicx} 
\usepackage{bm}

\begin{document}

\title{$J_{\rm eff}$=$\frac{1}{2}$ Mott spin-orbit insulating state close to the cubic limit in Ca$_4$IrO$_6$}

\author{S.~Calder}
\email{caldersa@ornl.gov}
\affiliation{Quantum Condensed Matter Division, Oak Ridge National Laboratory, Oak Ridge, TN 37831.}

\author{G.-X.~Cao}
\affiliation{Department of Materials Science and Engineering, University of Tennessee, Knoxville, TN 37996.}
\affiliation{Materials Science and Technology Division, Oak Ridge National Laboratory, Oak Ridge, TN 37831.}

\author{S.~Okamoto}
\affiliation{Materials Science and Technology Division, Oak Ridge National Laboratory, Oak Ridge, TN 37831.}

\author{J. W.~Kim}
\affiliation{Advanced Photon Source, Argonne National Laboratory, Argonne, IL 60439.}

\author{V.~R.~Cooper}
\affiliation{Materials Science and Technology Division, Oak Ridge National Laboratory, Oak Ridge, TN 37831.}

\author{Z.~Gai}
\affiliation{Center for Nanophase Materials Sciences, Oak Ridge National Laboratory, Oak Ridge, TN 37830.}

\author{B.~C.~Sales}
\affiliation{Materials Science and Technology Division, Oak Ridge National Laboratory, Oak Ridge, TN 37831.}

\author{M.~D.~Lumsden}
\affiliation{Quantum Condensed Matter Division, Oak Ridge National Laboratory, Oak Ridge, TN 37831.}

\author{D.~Mandrus}
\affiliation{Department of Materials Science and Engineering, University of Tennessee, Knoxville, TN 37996.}
\affiliation{Materials Science and Technology Division, Oak Ridge National Laboratory, Oak Ridge, TN 37831.}
 
\author{A.~D.~Christianson}
\affiliation{Quantum Condensed Matter Division, Oak Ridge National Laboratory, Oak Ridge, TN 37831.}

\begin{abstract}
The $J_{\rm eff}$=$\frac{1}{2}$ state is manifested in systems with large cubic crystal field splitting and spin-orbit coupling that are comparable to the on-site Coulomb interaction, $U$.  5d transition metal oxides host  parameters in this regime and strong evidence for this state in Sr$_2$IrO$_4$, and additional iridates, has been presented. All the candidates, however, deviate from the cubic crystal field required to provide an unmixed canonical $J_{\rm eff}$=$\frac{1}{2}$ state, impacting the development of a robust model of this novel insulating and magnetic state.  We present results that not only show Ca$_4$IrO$_6$ hosts the state, but furthermore uniquely resides close to the ideal case required for an unmixed $J_{\rm eff}$=$\frac{1}{2}$ state. 
 \end{abstract}

\maketitle

The competition between spin-orbit coupling (SOC), on-site Coulomb interaction and crystalline electric field (CEF) splitting in certain 5d iridates results in an electronic configuration that supports a novel $J_{\rm eff}$=$\frac{1}{2}$ Mott spin-orbit insulating state \cite{KimScience}. This state has been observed in Sr$_2$IrO$_4$ (Sr-214) \cite{KimScience}, Sr$_3$Ir$_2$O$_7$ (Sr-327) \cite{PhysRevLett.109.037204, PhysRevB.85.184432}, Ba$_2$IrO$_4$ (Ba-214) \cite{PhysRevLett.110.117207}, CaIrO$_3$ \cite{PhysRevLett.110.217212}, BaIrO$_3$ \cite{PhysRevLett.105.216407}, Na$_2$IrO$_3$ \cite{PhysRevLett.108.127203} and has been shown to remain upon dilution of the Ir ion site with, for example, Mn-substitution in Sr-214 \cite{CalderMnSr214}.  Common to all candidate $J_{\rm eff}$=$\frac{1}{2}$ materials is the coexistence of magnetic order within the insulating state, with the state remaining robust in alternative magnetic structures. Similarities to the high T$_c$ superconducting cuprates in  Sr-214 have added an extra dimension of interest \cite{PhysRevLett.108.177003}.  Open questions, however, have emerged regarding the role of crystal symmetry, SOC and magnetism, as well as debate over whether unequivocal experimental evidence of the $J_{\rm eff}$=$\frac{1}{2}$ state  exists \cite{0953-8984-23-25-252201}.

The $J_{\rm eff}$=$\frac{1}{2}$ state requires local cubic symmetry resulting in isotropic CEF splitting, in addition to appreciably strong SOC \cite{AbragamBleaney}. Significantly all proposed $J_{\rm eff}$=$\frac{1}{2}$ candidates, however, deviate from the cubic limit resulting in a mixed $J_{\rm eff}$=$\frac{1}{2}$,$\frac{3}{2}$  ground state. For example in Sr-214 the octahedra are elongated by $\sim$$4\%$ along the c-axis resulting in tetragonal corrections to the ground state \cite{PhysRevLett.102.017205}.  A consideration of Sr$_3$CuIrO$_6$, with a  non-cubic deviation of the O-Ir-O octahedra angle from 90$^{\circ}$ to 80$^{\circ}$, proposed that the result was a mixing away from a  pure  $J_{\rm eff}$=$\frac{1}{2}$ ground state due to the anisotropic CEF splitting \cite{PhysRevLett.109.157401}. Additionally, debate has emerged as to what role, if any, the magnetic order that occurs in the insulating phase plays. Indeed there has been suggestions that rather than a Mott insulating state, that necessitates a Coulomb driven opening of the band gap, a magnetically driven opening of the band gap via the Slater mechanism plays a crucial role \cite{AritaSlater}.  Finding a material closer to the ideal cubic case should aid the building of a robust model of the insulating and magnetic state in $J_{\rm eff}$=$\frac{1}{2}$ iridates. 
 
We consider such a candidate material, Ca$_4$IrO$_6$  (space group $R\bar3c$), an insulator at room temperature that contains Ir$^{4+}$ ions, as found in previous $J_{\rm eff}$=$\frac{1}{2}$ insulators \cite{sarkozy,DavisCairo,SegalCairo}. Initial investigations of Ca$_4$IrO$_6$ considered the Ir ions  as residing on quasi 1-D spin chains on a frustrated lattice \cite{DavisCairo}, a topology that could be expected to suppress long range magnetic order. Subsequent specific heat and susceptibility measurements \cite{SegalCairo, PhysRevB.75.134402} and a recent $\mu$SR investigation \cite{PhysRevB.83.094416}, however, indicated the onset of magnetic correlations around 12 K. In Ca$_4$IrO$_6$ the IrO$_6$ octahedra are well separated and disconnected, with no shared Ir-O-Ir bonds between octahedra. The octahedra contain identical Ir-O bond lengths, with a deviation of the O-Ir-O angle of  $<$2$^{\circ}$ from 90$^{\circ}$. Subsequently the IrO$_6$ octahedral environment  in Ca$_4$IrO$_6$ resides close to the ideal cubic limit required for the CEF and SOC splitting of the d-manifold to produce an unmixed $J_{\rm eff}$=$\frac{1}{2}$.  We present neutron, RXS and density functional theory results that support this postulate and additionally report the development of long range magnetic order.  

Single crystals of Ca$_4$IrO$_6$ with dimensions 0.1mm$^3$ were grown using the flux method. 4 grams of powder Ca$_4$IrO$_6$ was prepared using solid state techniques. Neutron scattering was performed on the powder sample at the High Flux Isotope Reactor (HFIR) using the triple axis instruments HB-3 and  HB-1A in elastic mode with a wavelength of 2.36 $\rm \AA$ and on the powder diffractometer HB-2A with wavelength 1.54 $\rm \AA$. An annular aluminum sample holder was utilized to reduce neutron absorption. Single crystal measurements were performed using magnetic resonant x-ray scattering (RXS) at the Advanced Photon Source (APS) on beamline 6-ID-B at both the $L_2$ (12.83 keV) and $L_3$ (11.22 keV) resonant edges of iridium. Graphite was used as the polarization analyzer at the (0 0 10) and (008) reflections on the $L_2$ and $L_3$ edges, respectively, to achieve a scattering angle close to 90$^\circ$. We performed polarization analysis of the photon beam in $\sigma$-$\pi$  and $\sigma$-$\sigma$ mode to separate out the magnetic and charge scattering, respectively. To observe the sample fluorescence, energy scans were performed without the analyzer and with the detector away from any Bragg peaks through both absorption energies. The DFT calculations were performed with the generalized gradient approximation (GGA) and projector augmented wave (PAW) approach \cite{PhysRevB.50.17953} as implemented in the Vienna Ab initio Simulation Package (VASP) \cite{PhysRevB.54.11169,PhysRevB.59.1758}, with relativistic SOC included. The experimental structure in Ref.~\onlinecite{SegalCairo} was used with a 2$\times$2$\times$2 k-point grid, doubled unit cell and an energy cutoff of 550 eV. We employ the rotationally invariant LDA+U method of Liechtenstein \cite{PhysRevB.52.R5467} with U=2 and J=0.2 eV for the Ir d states, values consistent with the literature \cite{PhysRevLett.101.076402, AritaSlater}. For Ir and O, we use standard potentials (Ir and O in the VASP distribution), and for Ca, we use PAW potentials in which semicore s and p states are treated as valence states (Capv).

\begin{figure}[bt]
   \centering         
      \includegraphics[width=0.9\columnwidth]{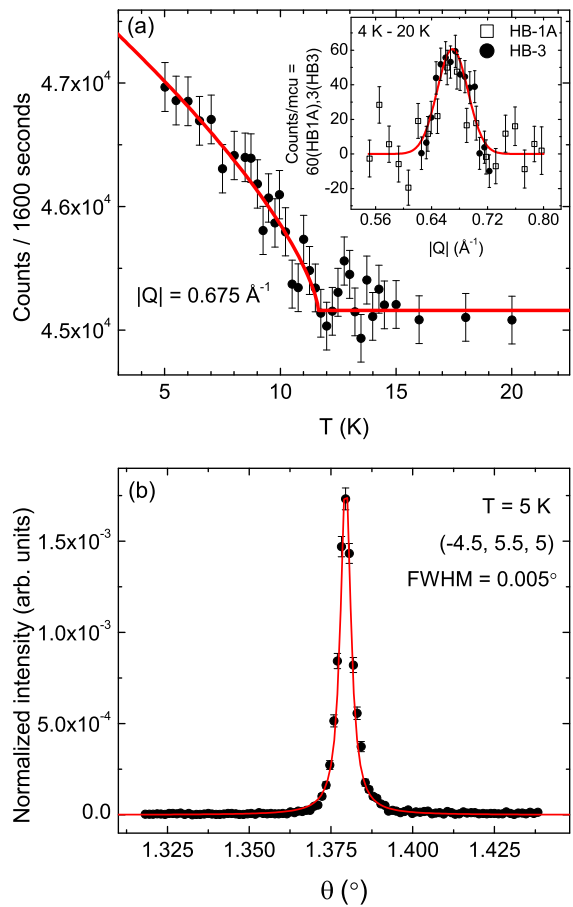}  
                       \caption{\label{FigureNeutronandXray} (a) Elastic neutron diffraction on beamlines HB-3 and HB-1A around $|Q|$=0.67 $\rm \AA^{-1}$. (Inset) 4 K minus 20 K data. The results are scaled to account for the different neutron flux on each instrument. (Main panel) Intensity measured at $|Q|$=0.67 $\rm \AA^{-1}$ using HB-3 showing the onset of magnetic ordering around 12 K. (b) RXS measurement of the magnetic (-4.5,5.5,5) reflection of Ca$_4$IrO$_6$ at the $L_3$ resonant edge (11.215 keV) in $\sigma$-$\pi$ mode.}
\end{figure}

We begin our investigation with a consideration of the magnetic ordering in Ca$_4$IrO$_6$, having first ruled out any thermal structural anomalies through a neutron powder diffraction investigation. Low $|\rm Q|$ neutron measurements, see Fig.~\ref{FigureNeutronandXray}, show increased scattering intensity, indicative of magnetic order, at $|\rm Q|$=0.67 $\rm \AA^{-1}$ in going from 20 K to 4 K. The scattering is commensurate with the nuclear structure, but at a non-integer wavevector, making it compatible with antiferromagnetic ordering. Measurements on both HB-3 and HB-1A suggested a magnetic reflection at $|\rm Q|$=1.17 $\rm \AA^{-1}$, as indicated by increased intensity below 12 K. The intensity variation of the scattering at $|\rm Q|$=0.67 $\rm \AA^{-1}$ with temperature is consistent with a second order transition, as shown in Fig.~\ref{FigureNeutronandXray}. Fitting the results to a power law gives a transition temperature of 12 K.

To explore the ordered magnetic structure we performed a representational analysis \cite{sarahwills}. The propagation vector ${\bf k}$$=$($\frac{1}{2} \frac{1}{2} 0$) produces magnetic reflections at both $|\rm Q|$=0.67 $\rm \AA^{-1}$ and $|\rm Q|$=1.17 $\rm \AA^{-1}$  and no other low angle positions, consistent with the neutron data and as we show single crystal x-ray data.  Alternative propagation vectors were considered, but all failed to give intensity at $|\rm Q|$=0.67 $\rm \AA^{-1}$. The space group $R\bar3c$ and Ir ions on the $6b$ Wyckoff position gives the irreducible representations (IR) $\Gamma_1$ and $\Gamma_3$ (following the numbering scheme of Kareps). We employ the simplification for a second order transition that only one IR describes the magnetic structure \cite{IRcomment}. Our neutron results alone do not allow a unique determination of the magnetic structure. 

RXS was performed on a single crystal of Ca$_4$IrO$_6$ with a tight mosaic ($\sim$0.005$^{\circ}$), see Fig.~\ref{FigureNeutronandXray}. We surveyed a large number of reflections and performed a polarization analysis to confirm the magnetic  or non-magnetic nature of each reflection, observing 30 magnetic reflections. All magnetic peaks were consistent with the ${\bf k}$$=$($\frac{1}{2} \frac{1}{2} 0$) propagation vector. The predicted intensity contributed from each basis vector within the two symmetry permitted IRs allows the possible moment directions to be considered. For $\Gamma_{1}$ ($\Gamma_{3}$) with moments in the $ab$-plane only even (odd) L reflections are allowed, conversely for moments along the c-axis only odd (even) L reflections are allowed. We observed both even and odd L reflections indicating the spins are not confined to either the $ab$-plane or $c$-axis, independent of whether the magnetic structure is described by  $\Gamma_{1}$ or $\Gamma_{3}$.  This spin direction constrain produces the same allowed reflections for both $\Gamma_{1}$ or $\Gamma_{3}$.

\begin{figure}[t]
   \centering                   
      \includegraphics[trim=1.3cm 2.1cm 1.3cm 6.1cm,clip=true, width=1.0\columnwidth]{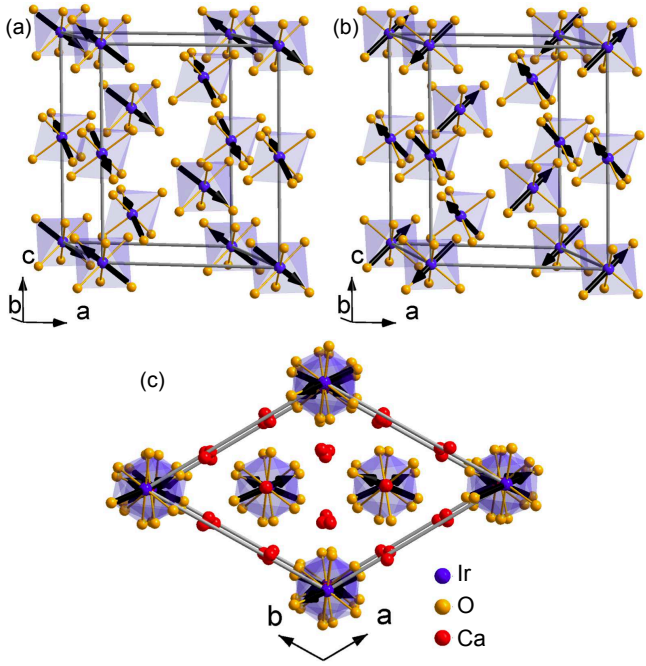} 
                   \caption{\label{FigXtalStructure} Candidate magnetic structures for Ca$_4$IrO$_6$ for the (a) $\Gamma_1$ and (b) $\Gamma_3$ IRs based on experimental results and DFT calculations. Only the nuclear unit cell is shown, with $a$=9.3 $\rm \AA$, $c$=11.2 $\rm \AA$. The magnetic unit cell is doubled in the $a$-$b$ plane.  The lowest energy magnetic structures have the spin directions along the Ir-O bond in the octahedra. (c) $a$-$b$ plane view of the $\Gamma_1$ magnetic structure showing the separated IrO$_6$ octahedra.}
\end{figure}

To allow for a distinction between the two candidate magnetic structures, we performed DFT calculations using VASP.  Starting with the spins in the $a$-$c$ plane in both $\Gamma_{1}$ and $\Gamma_{3}$ configurations, the spins were unconstrained and stable magnetic solutions were obtained for both cases, shown in Fig. \ref{FigXtalStructure}. We note that this is distinct from the case with the spins confined to the $a$-$b$ plane, where stable magnetic solutions could only be obtained by  applying extra constraints and fixing the spins. For $\Gamma_{1}$, the direction of magnetic moments rotated from the $a$-$c$ plane to the $b$-$c$ plane with the original symmetry. For both candidate magnetic structures the spins where initially not aligned along the Ir-O bond directions in the octahedra, however they relaxed along the Ir-O directions to form the lowest energy ground state. This is similar to the behavior observed in Sr-214, where the octahedra rotation controls the spin directions and may be a general feature of magnetically ordered $J_{\rm eff}$=$\frac{1}{2}$ iridates \cite{PhysRevLett.102.017205}.  For Ca$_4$IrO$_6$ the $\Gamma_{1}$ magnetic structure is lower in energy, compared to the $\Gamma_{3}$ magnetic structure, by $\sim$0.018eV/unit cell, $\sim$0.0015eV/site. With $\rm T_N$$=$12 K this is an appreciably different energy scale. We therefore conclude that $\Gamma_{1}$, with the spins aligned as shown in Fig.~\ref{FigXtalStructure}(a), is the most likely long ranged magnetic structure for Ca$_4$IrO$_6$. Applying our DFT results for the magnetic structure to the experimental neutron data, and normalizing the intensity with respect to the nuclear reflections, gives an ordered magnetic moment of 0.42(10)$\mu_B$. This value corresponds to, within experimental error, the ordered moments for the $J_{\rm eff}$=$\frac{1}{2}$ materials Sr-214 (0.208(3)-0.36(6)$\mu_B$)\cite{PhysRevB.87.144405,PhysRevB.87.140406}, Sr-327 (0.35(6)$\mu_B$)\cite{PhysRevB.86.100401} and Mn-doped Sr-214 (0.5(1)$\mu_B$)\cite{CalderMnSr214} found from  neutron investigations.

\begin{figure}[t]
   \centering                   
      \includegraphics[width=1.0\columnwidth]{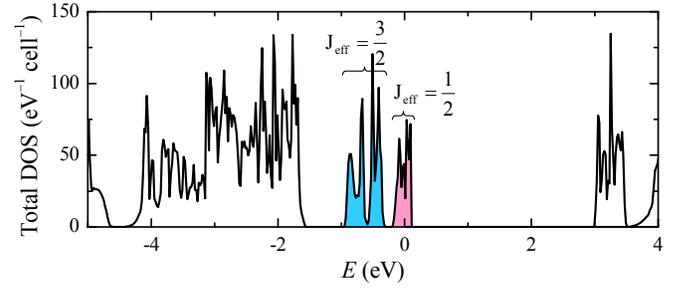} 
                   \caption{\label{FigDOS} Total density of states (DOS) of Ca$_4$IrO$_6$ for the non-magnetic case, with the inclusion of SOC and U.}
\end{figure}

\begin{figure}[tb]
   \centering                 
                       \includegraphics[trim=0cm 3cm 0cm 3cm,clip=true, width=1.0\columnwidth]{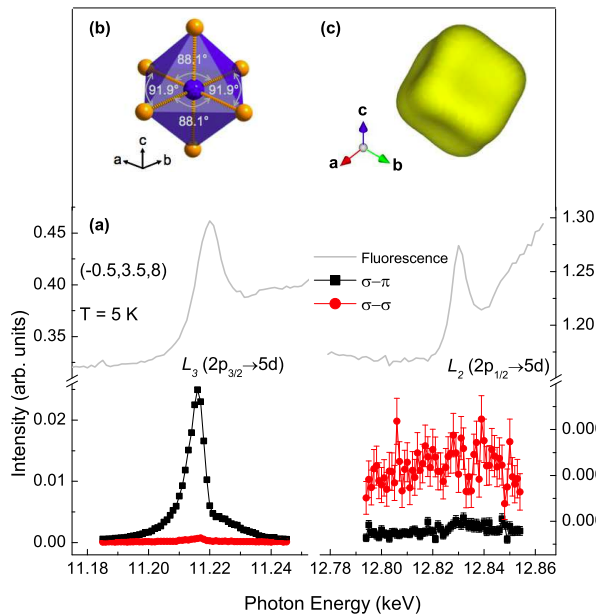} 
                          \caption{\label{FigureL2andL3edges} (a) RXS energy scans through the  $L_3$ (11.22 keV) and $L_2$  (12.83 keV)  Ir edges at the (-0.5,3.5,8) magnetic reflection in Ca$_4$IrO$_6$.  A large resonant enhancement occurs at $L_3$, whereas the intensity is suppressed at the $L_2$ energy. This behavior has been reported to be a signature of the  $J_{\rm eff}$=$\frac{1}{2}$  state. (b) Local environment of the Ir ions probed by RXS, the IrO$_6$ environment deviates by  $<$$2\%$ from the ideal case. (c) Total charge density plot around an Ir ion in Ca$_4$IrO$_6$ shows a highly symmetric charge distribution.}
\end{figure}

We now turn to the insulating state of Ca$_4$IrO$_6$. DFT calculations, shown in Fig.~\ref{FigDOS}, indicate that SOC, along with the inherent CEF in Ca$_4$IrO$_6$, breaks the $t_{2g}$ manifold degeneracy into  $J_{\rm eff}$=$\frac{1}{2}$ and $J_{\rm eff}$=$\frac{3}{2}$ states. The $J_{\rm eff}$=$\frac{1}{2}$ band width in Ca$_4$IrO$_6$ (about 0.2eV) is much narrower than that in Sr$_2$IrO$_4$ (about 1eV). Consequently even a moderate U is expected to induce a Mott insulating state once one goes beyond the DFT+U calculations, for example DFT+DMFT. For DFT+U the introduction of magnetic order splits the $J_{\rm eff}$=$\frac{1}{2}$ band, resulting in insulating states with a gap amplitude about 0.6eV for both $\Gamma_1$ and $\Gamma_3$ structures.   Experimentally, a comparison of the intensity enhancement at the $L_2$ and $L_3$ edges in RXS measurements has been shown to uncover a signature of the $J_{\rm eff}$=$\frac{1}{2}$ state, initially in Sr-214 and then subsequently in Sr-327 and Ba-214 \cite{KimScience, PhysRevB.85.184432,PhysRevLett.110.117207}. Specifically, for a 10$Dq$ CEF split d-manifold in the limit of negligible SOC splitting for Ir$^{4+}$, i.e. the case of S=1/2, the $t_{2g}$ manifold with both J=3/2 and J=5/2 is the lowest unoccupied state. Consequently both the $L_2$-edge ($2p_{\frac{1}{2}} \rightarrow 5d_{\frac{3}{2}}$) and $L_3$-edge ($2p_{\frac{3}{2}} \rightarrow 5d_{\frac{3}{2}},_{\frac{5}{2}}$) RXS transitions are allowed and appreciable intensity is expected at both edges during a  measurement. Conversely for large SOC and CEF splitting, i.e. the $J_{\rm eff}$=$\frac{1}{2}$ case, the $J_{\rm eff}$=$\frac{3}{2}$ band in Ir$^{4+}$ produced from the J = 3/2 states will be fully occupied forbidding a transition at the $L_2$ edge, resulting in no measured RXS intensity. Measurements at the magnetic reflection (-0.5, 3.5, 8) through both  $L_2$ and $L_3$ resonant energies in Ca$_4$IrO$_6$ are shown in Fig \ref{FigureL2andL3edges}(a).  A large enhancement is observed at the $L_3$ edge that has its maximum at the inflection point of the fluorescence scans, as expected. Conversely there is no appreciable resonant enhancement at the $L_2$ energy, with the fluorescence scan showing a maximum at the expected $L_2$ edge energy. Following the reasoning in Ref.~\onlinecite{KimScience} this is direct evidence for a $J_{\rm eff}$=$\frac{1}{2}$ state in Ca$_4$IrO$_6$.   

All previous candidate $J_{\rm eff}$=$\frac{1}{2}$ materials diverge from a local cubic IrO$_6$ environment.  Such deviations have been suggested to control the ratio of the resonant intensity I$_L{_2}$/I$_L{_3}$ \cite{PhysRevLett.102.017205,PhysRevLett.109.037204}. The ideal cubic case yields an unmixed $J_{\rm eff}$=$\frac{1}{2}$ state and subsequently the I$_L{_2}$/I$_L{_3}$ ratio should be identically zero. Previous RXS investigations of $J_{\rm eff}$=$\frac{1}{2}$ materials Sr-214, Sr-327 and Ba-214 have involved Ir octahedra with axial distortions, quantified by a tetragonal splitting variable $\Delta$. Consequently a factor $\theta$, where $\tan(2\theta) = 2\sqrt 2 \lambda / (\lambda-2\Delta) $, with $\lambda$ the SOC, has been introduced to parameterize the tetragonal distortion with SOC and account for observed deviations from the ideal I$_L{_2}$/I$_L{_3}$ ratio \cite{PhysRevLett.102.017205}.  $\Delta$=$0$ for Ca$_4$IrO$_6$ and as reported in Ref.~\onlinecite{PhysRevLett.109.037204} this should be manifested in I$_L{_2}$/I$_L{_3}$=0 and a symmetric spin-density profile map. Experimentally for Ca$_4$IrO$_6$ we find no appreciable intensity at the $L_2$ edge, consistent with isotropic CEF splitting and strong SOC creating a $J_{\rm eff}$=$\frac{1}{2}$ ground state. In order to quantify this difference in intensity we find a ratio I$_L{_2}$/I$_L{_3}$=3.2(6)$\times$10$^{-4}$, significantly smaller than our estimates of I$_L{_2}$/I$_L{_3}$ for Sr-214 and Sr-327 \cite{Ratiocomment}. Furthermore DFT calculations of the total charge density around the Ir ion, including SOC, show local cubic symmetry as evidenced by the dimpled dice shape in Fig. \ref{FigureL2andL3edges}(c) and the total DOS in Fig.~\ref{FigDOS} reveals well separated $J_{\rm eff}$=$\frac{1}{2}$ and $J_{\rm eff}$=$\frac{3}{2}$ bands. Therefore Ca$_4$IrO$_6$ appears to reside close to the ideal cubic and strong SOC limit required for an unmixed $J_{\rm eff}$=$\frac{1}{2}$ ground state.

Contrasting further with Sr-214 is the low $\rm T_N$ of Ca$_4$IrO$_6$. This may be driven by the geometric frustration inherent in the crystal structure, as indicated by the large ratio of $\Theta_{\rm CW}$/T$\rm _N\approx$ 3-5 \cite{PhysRevB.75.134402}.  While debate exists in Sr-214 as to whether the magnetically driven Slater mechanism plays a significant role in creating a band gap and driving the insulating state, the large separation between the insulating state ($>$ 300 K) and T$\rm _N$ (12 K) suggests a purely Mott insulating state in Ca$_4$IrO$_6$.  

Our experimental and theoretical results have shown that Ca$_4$IrO$_6$ is a new candidate $J_{\rm eff}$=$\frac{1}{2}$ Mott spin-orbit insulator, with long range magnetic order occurring well within the insulating phase. While previous candidate materials have fallen outside the limit of cubic crystal field splitting, resulting in a mixing away from a pure $J_{\rm eff}$=$\frac{1}{2}$ ground state, Ca$_4$IrO$_6$ resides close to the limit required for an unmixed ground state. Consequently Ca$_4$IrO$_6$ is uniquely positioned to act as a canonical example of a $J_{\rm eff}$=$\frac{1}{2}$ system.

This research at ORNL's High Flux Isotope Reactor was sponsored by the Scientific User Facilities Division, Office of Basic Energy Sciences, U.S. Department of Energy. Part of the work (DM, BCS, GC and SO) was supported by the Department of Energy, Basic Energy Sciences, Materials Sciences and Engineering Division. Use of the Advanced Photon Source, an Office of Science User Facility operated for the U.S. DOE Office of Science by Argonne National Laboratory, was supported by the U.S. DOE under Contract No. DE-AC02-06CH11357.


\begin{thebibliography}{30}
\bibitem [{IRc()}]{IRcomment}%
  \BibitemOpen
  \href@noop {} {}\bibinfo {note} {We treat the magnetic transition in
  Ca$_4$IrO$_6$ as continuous based on our experimental results and previous
  measurements in the literature in Refs.
  \cite{PhysRevB.75.134402,PhysRevB.83.094416}. The ratio of the number of
  symmetry operations through the transition N(G$_0$)/n(G$_k$)=3, this may
  indicate a weakly first order transition \cite{Cracknell}.}\BibitemShut
  {Stop}%
\bibitem [{Rat()}]{Ratiocomment}%
  \BibitemOpen
  \href@noop {} {}\bibinfo {note} {We considered the resonant edge intensities
  from previous published candidate materials to obtain estimates of
  I$_L{_2}$/I$_L{_3}$ values, while noting the difficulty in quantitatively
  comparing different experiments and materials using RXS. We find from Ref.
  \onlinecite{KimScience} Sr-214 has a ratio of 4.3$\times$10$^{-2}$ and for
  Sr-327 ratios of 9.6$\times$10$^{-3}$ and 1.4$\times$10$^{-2}$ from Refs.
  \onlinecite{PhysRevLett.109.037204, PhysRevB.85.184432}.}\BibitemShut {Stop}%
\end{thebibliography}
\end{document}